\documentclass[final,authoryear,5p]{elsarticle}

\usepackage{graphicx}

\usepackage{amssymb}
\usepackage{amsthm}
\usepackage{amsmath}
\usepackage{tabularx}

\numberwithin{table}{section}

\usepackage{hyperref}

\journal{Advances in Space Research}

\begin{document}

\begin{frontmatter}
\title{Design and performance of a vacuum-UV simulator for material testing under space conditions}


\author[uhb]{M. Sznajder}

\author[dlr]{T. Renger\corref{cor}}
\cortext[cor]{Corresponding author}
\ead{Thomas.Renger@dlr.de}

\author[dlr]{A. Witzke}

\author[dlr,zgu]{U. Geppert}

\author[ptb]{R. Thornagel}

\address[uhb]{DLR Institute for Space Systems, University of Bremen, Robert Hooke Str. 7, 28359 Bremen, Germany}
\address[dlr]{DLR Institute for Space Systems, System Conditioning, Robert Hooke Str. 7, 28359 Bremen, Germany}
\address[zgu]{Kepler Institute of Astronomy, University of Zielona G\'{o}ra, 65-265 Zielona G\'{o}ra, Lubuska 2, Poland}
\address[ptb]{Physikalisch-Technische Bundesanstalt, Abbestr. 2-12, 10587 Berlin, Germany}

\begin{abstract}
This paper describes the construction and performance of a VUV-simulator that has been designed to study degradation of materials under space conditions. It is part of the Complex Irradiation Facility at DLR in Bremen, Germany, that has been built for testing of material under irradiation in the complete UV-range as well as under proton and electron irradiation. Presently available UV-sources used for material tests do not allow the irradiation with wavelengths smaller than about $115$ nm where common Deuterium lamps show an intensity cut-off. The VUV-simulator generates radiation by excitation of a gas-flow with an electron beam. The intensity of the radiation can be varied by manipulating the gas-flow and/or the electron beam.

The VUV simulator has been calibrated at three different gas-flow settings in the range from $40$ nm to $410$ nm. The calibration has been made by the Physikalisch-Technische Bundesanstalt (PTB) in Berlin. The measured spectra show total irradiance intensities from $24$ to $58$ mW$\rm{m^{-2}}$ (see Table \ref{tab:long}) in the VUV-range, i.e. for wavelengths smaller than $200$ nm. They exhibit a large number of spectral lines generated either by the gas-flow constituents or by metal atoms in the residual gas which come from metals used in the source construction. In the range from $40$ nm to $120$ nm where Deuterium lamps are not usable, acceleration factors of $3$ to $26.3$ Solar Constants are reached depending on the gas-flow setting. The VUV-simulator allows studies of general degradation effects caused by photoionization and photodissociation as well as accelerated degradation tests by use of intensities that are significantly higher compared to that of the Sun at $1$ AU.  
\end{abstract}


\end{frontmatter}

\parindent=0.5 cm

\section{Introduction}
\label{intro}

All materials planned for space applications in which they will be exposed to the radiation in space have to be evaluated for their behavior under particle and UV radiation \citep{e512, ecss}. It is known from many of these evaluation tests that particle and UV radiation can significantly degrade materials and, e.g. lead to changes in their mechanical behavior or thermo-optical properties \citep{agno,heltzel,lura1,onera1,romero,sharma}. These changes can cause early failures of satellite components or even failures of complete space missions \citep{dach05}. 

\begin{figure*}
  \begin{center}
    \includegraphics*[width=0.8\textwidth]{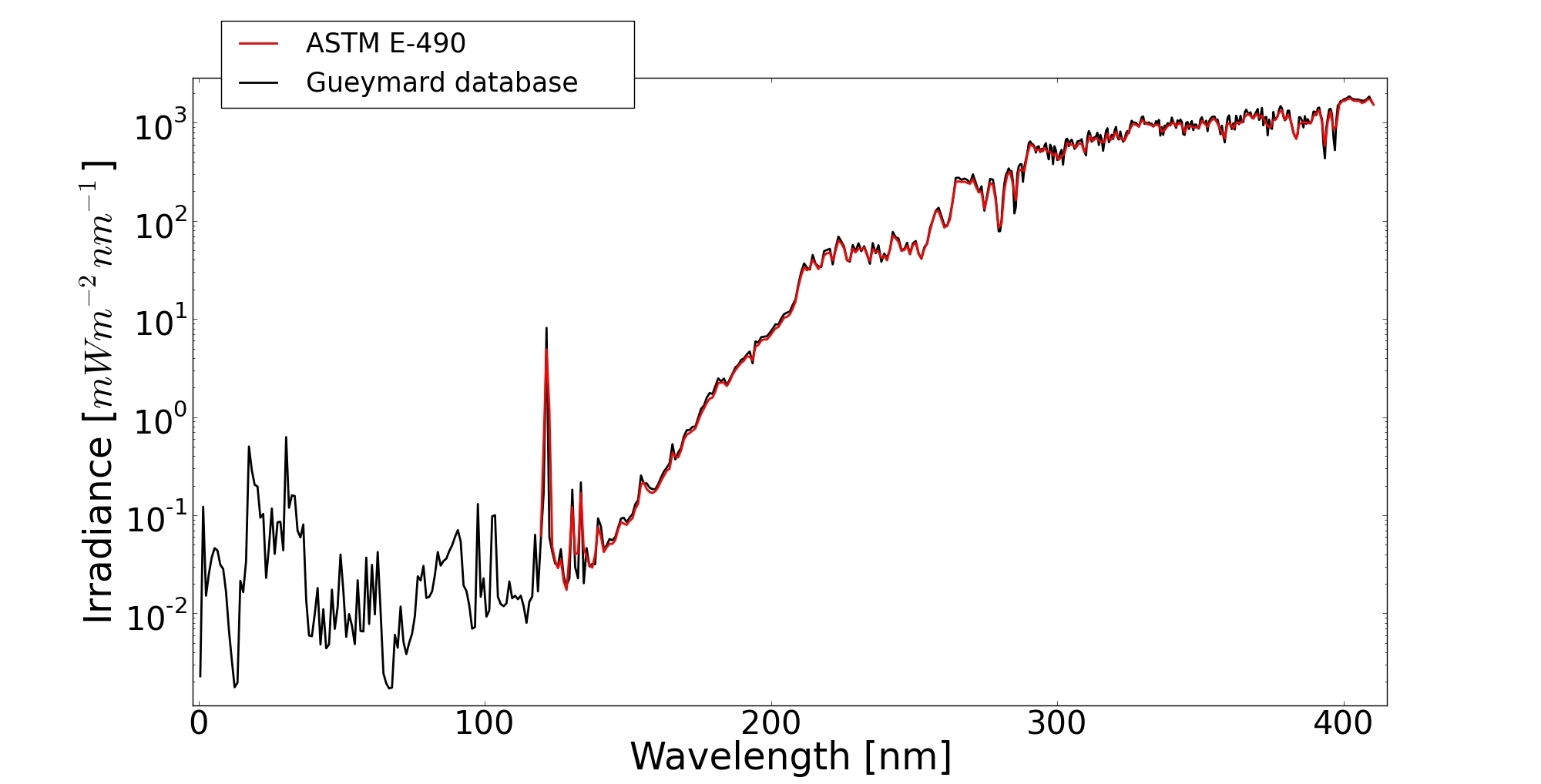}
  \end{center}
  \caption{Extraterrestrial spectral solar UV-irradiance at 1 AU (astronomic unit). The red line shows data from ASTM E-490 \citep{e490}. This standard provides data only down to $119.5$ nm. The black line exhibits spectral data from Gueymard \citep{guey}. This data are representative for average solar activity conditions. They are in very good accordance to the ASTM E-490 spectrum between $120$ nm and $400$ nm.}
  \label{standard}
\end{figure*}

This paper discusses the design and performance of a radiation source that has been developed to simulate the short wavelength part of the Vacuum-UV region of the solar spectrum during material tests. The solar UV radiation is generally defined as the solar radiation with wavelengths from $10$ nm to $400$ nm \citep{e512}. The range between $200$ nm and $400$ nm is named Near-UV (NUV) range. The other part of the solar UV radiation from $10$ nm to $200$ nm is denoted as the Vacuum-UV (VUV) range. Extraterrestrial intensity spectra of both ranges show that the contribution of the VUV radiation to the total intensity of solar UV irradiation is almost negligible (Fig. \ref{standard}). The VUV-irradiation amounts to only $0.1$\% of the NUV irradiation intensity. However, the VUV radiation can noticeably contribute to the degradation of materials despite of its small amount of total intensity. The energy of a single photon in the VUV range is considerably higher compared to a photon in the NUV range. VUV-photon energies vary from $6$ eV to $124$ eV whereas NUV-photon energies range only from $3$ eV to $6$ eV. Therefore, VUV radiation can generate photoionization and photodissociation effects that cannot be caused by the significantly lower photon energies in the NUV range. Thus, degradation effects that would not occur under NUV irradiation even at very high intensities may be expected if the material is exposed to VUV irradiation for longer periods of time. 

To get the most reliable information on UV degradation of materials in space by ground material tests, the UV-spectrum of the Sun should be simulated as close as possible. The NUV spectral region can be simulated by using commercial short arc Xenon lamps \citep{onera}. The VUV spectral range is simulated with Deuterium lamps in almost all material tests \citep[e.g.][]{polsak}. These lamps, however, feature a lower wavelength cut-off at approx. $115$ nm due to internal $\rm{MgF_2}$- or $\rm{LiF_2}$-windows which exhibit a strong decrease in transmission at this wavelength. This leads to the fact that the residual VUV wavelength range between $10$ nm and $115$ nm is generally not covered in up-to-date material tests although especially this range shows a strong increase of photon energies from $10$ eV to $124$ eV.

In context with the installation of the new Complex Irradiation Facility (CIF) for material testing at the German Aerospace Center (DLR) \citep{cif,cif2} which should be capable of simulating the complete solar UV irradiation as well as proton and electron irradiation, the lack in the VUV simulation of the solar spectrum by currently available radiation sources necessitated the construction of a VUV simulator that covers the range from $10$ nm to at least $115$ nm. Beside a good approximation to the real solar spectrum this simulator has to achieve several other requirements to make it useful for material testing: It has to exceed the radiation intensity of the Sun (at 1 AU) at the sample area of the CIF in order to accelerate the degradation of the tested materials and to allow the simulation of long-term effects. Furthermore, the simulator has to generate radiation that is emitted under a relatively large solid angle into the test chamber to permit the simultaneous irradiation of several samples with a homogenous intensity distribution. It must, moreover, be capable of working continuously during a sufficiently long period of time. The construction must take into account that the VUV simulator has to be connected to the test chamber without any window as no window material is known that completely transmits radiation in the concerned spectral range. Therefore, the design has to ensure that no significant amount of particles of the medium necessary to generate the radiation can migrate inside the test chamber.

Below we describe in detail the design and performance of a VUV simulator that has been built with regard to the given requirements above. It bases on a design of a VUV gas-jet simulator that was constructed $15$ years ago by the Institute of Low Temperature Physics and Engineering in Kharkov, Ukraine, in collaboration with the DLR \citep{ver}. The calibration of the VUV simulator has been carried out in the range from $40$ nm to $410$ nm by the Physikalisch Technische Bundesanstalt (PTB) in Berlin, Germany. The calibration method and procedure is briefly explained in Section \ref{calibration}. The spectral distribution of radiation as well as the irradiance, the derived acceleration factors as well as the stability of the source are discussed and summarized in the ensuing Sections \ref{spectra}, \ref{acc} and \ref{stability}. The summary is given in Section \ref{summary}.

\section{Design and Principle of Operation of the VUV-Source}
\label{vuv}

The design of the VUV simulator is based on the semi-cryogenic version of the previous simulators which are described in \citep{ver}. The radiation is produced by the transition of electrons belonging to excited gas atoms into their ground state. The gas atoms are excited by electron bombardment of a spatially limited supersonic gas jet which flows into a vacuum chamber. The vacuum is maintained by a combination of cryogenic and mechanical pumps (semi-cryogenic design). A pressure of about $1$ mbar inside the jet is sufficient for an effective excitation. Beyond the jet close to its boundary the gas pressure is several orders less ($10^{-4}$ mbar) caused by the supersonic directional motion of the jet. This is the premise to locate an electron source in the close vicinity of the gas jet. This method generates electromagnetic radiation in a broad spectral range (soft X-rays, VUV, NUV) with relatively high intensities at lower wavelengths ($<115$ nm) and permits a design without windows, which would disable the transmission in this range. The spectral intensity distribution of the radiation depends on the gas mixture, the flow rate of the jet, the electron current (on the alignment between beam and jet, and the focus of the electron beam).

\begin{figure*}[!t]
  \begin{center}
    \includegraphics*[width=0.8\textwidth]{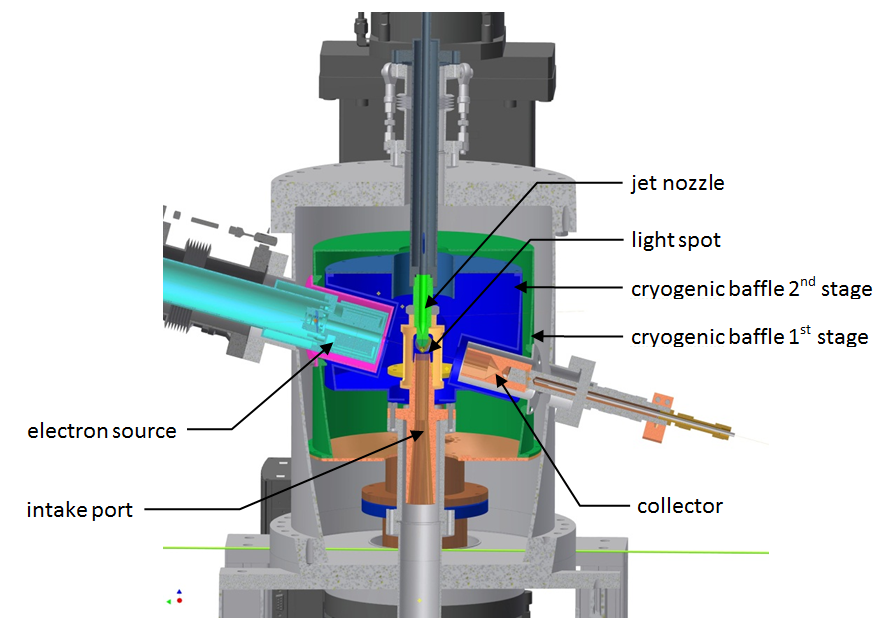}
  \end{center}
  \caption{Sectioning of the VUV-simulator along the electron beam.}
  \label{vuv_1}
\end{figure*}

\begin{figure*}[!t]
  \begin{center}
    \includegraphics*[width=0.8\textwidth]{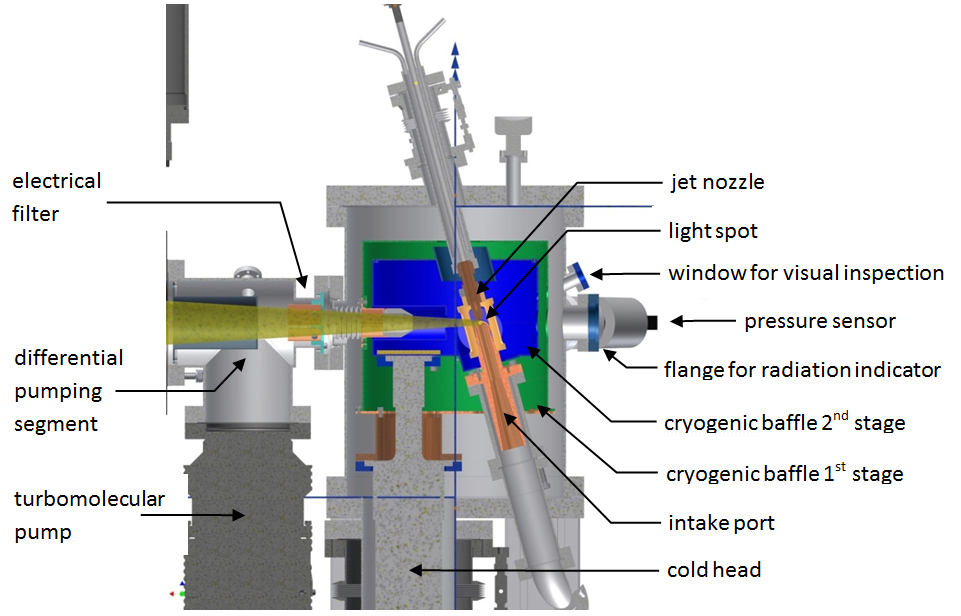}
  \end{center}
  \caption{Sectioning of the VUV-simulator along the light outlet (yellow).}
  \label{vuv_2}
\end{figure*}

Fig. \ref{vuv_1} illustrates the arrangement of the electron source and the gas jet inside the vacuum chamber of the simulator. The outlet of the generated VUV-light is realized behind the spot perpendicular to the figure’s plane. The electrons which pass through the gas jet are caught by the collector at the opposite side of the source. The components of the vacuum system are better visible in Fig. \ref{vuv_2}, whose plane is perpendicular to the plane of Fig. \ref{vuv_1}.

The gas jet is injected by a nozzle from top of the VUV-chamber into the vacuum. The flow rate is stabilized by a flow controller and is adjustable by software in the range between $0$ and $5000$ sccm (standard cubic centimeter per minute). The bulk of the gas load is pumped out through an intake port at the bottom of the chamber by a screw pump. The rest, which is a small fraction of about $3$\% is removed by cryogenic condensation at two baffles. Each of them is connected to one of the both stages of the cold head of a commercial cryogenic pump (Helium cooling machine). The $2^{\rm{nd}}$ stage reaches a bottom temperature of $15$ K without gas load. It increases to about $20$ K under gas load by formation of ice, which decreases the pumping power gradually and limits the operating time. The temperature is logged permanently as an indicator when a regeneration is necessary to remove the ice. During this process the cryogenic pump is turned off while the mechanical pumps continue their operation. The temperature inside the chamber increases and the residual gas, which is frozen on the cold baffles, is pumped out. The gas nozzle is thermally connected to a water circuit for tempering it to avoid a freeze. The combination of both pumping procedures is possible due to the effect that the gas inside the intake port banks itself. Thereby, a reverse flow from the pipeline is impeded and the pressure regions are separated by $10^{-4}$ mbar around the gas jet (inside the chamber) and more than $10^{-2}$ mbar inside the pipeline. 

A differential pumping segment is installed at the light outlet consisting of a turbomolecular pump and an aperture assembly. It improves the pressure conditions inside the radiation chamber and reduces the metal traces to the lowest possible limit. Since tests can be performed in the chamber at pressures of $10^{-6}$ mbar and less the pollution by metals over the test period is expected to be negligible. The fraction of light which is permitted to the test chamber is colored yellow in Fig. \ref{vuv_2}. An opening angle of about $6^{\rm{o}}$ is defined by the apertures in the cryogenic baffles and at the differential pumping segment. It ensures the irradiation of a target area with a diameter of $80$ mm at a distance of about $770$ mm from the spot, as it is realized in CIF. To avoid that charged particles produced at the spot can reach the irradiated object, an electric filter is installed at the light outlet, which deflects them beyond the radiation flow.

The axes of the gas jet, the electron beam and the light outlet are arranged out of square, while the orientation of the light outlet is horizontal. The axis of the gas jet is turned by $15^{\rm{o}}$ to the vertical axis in the direction of the light outlet. The electron beam is inclined by $15^{\rm{o}}$ to the horizontal axis but in its vertical plane it is perpendicular to the light outlet. The idea of this design is that the probability that particles could reach the radiation chamber or damage the cathode of the electron source is less than for a perpendicular arrangement.

The electron source is realized as a Pierce-type model with a magnetic lens behind the anode. The La$\rm{B_{6}}$-cathode is heated up electrically by adjusting the cathode voltage depending on which emission current is needed. The electronic control unit of the source provides a PID algorithm which stabilizes the emission current by varying the Wehnelt voltage automatically. The beam is focused by setting the current for the magnetic lens.

Two flanges are located at the opposite side of the light outlet. Each is connected with a window for visual inspection of the luminous jet and with a radiation indicator compartment (see Fig. \ref{vuv_2}). The indicator measures the signal of the source. The digital value of the signal is visualized by the controlling software to monitor the stability of the radiation. During the calibration procedure for every measured spectrum the corresponding sensor value has been determined.

\begin{figure*}[!t]
  \begin{center}
    \includegraphics*[width=0.6\textwidth]{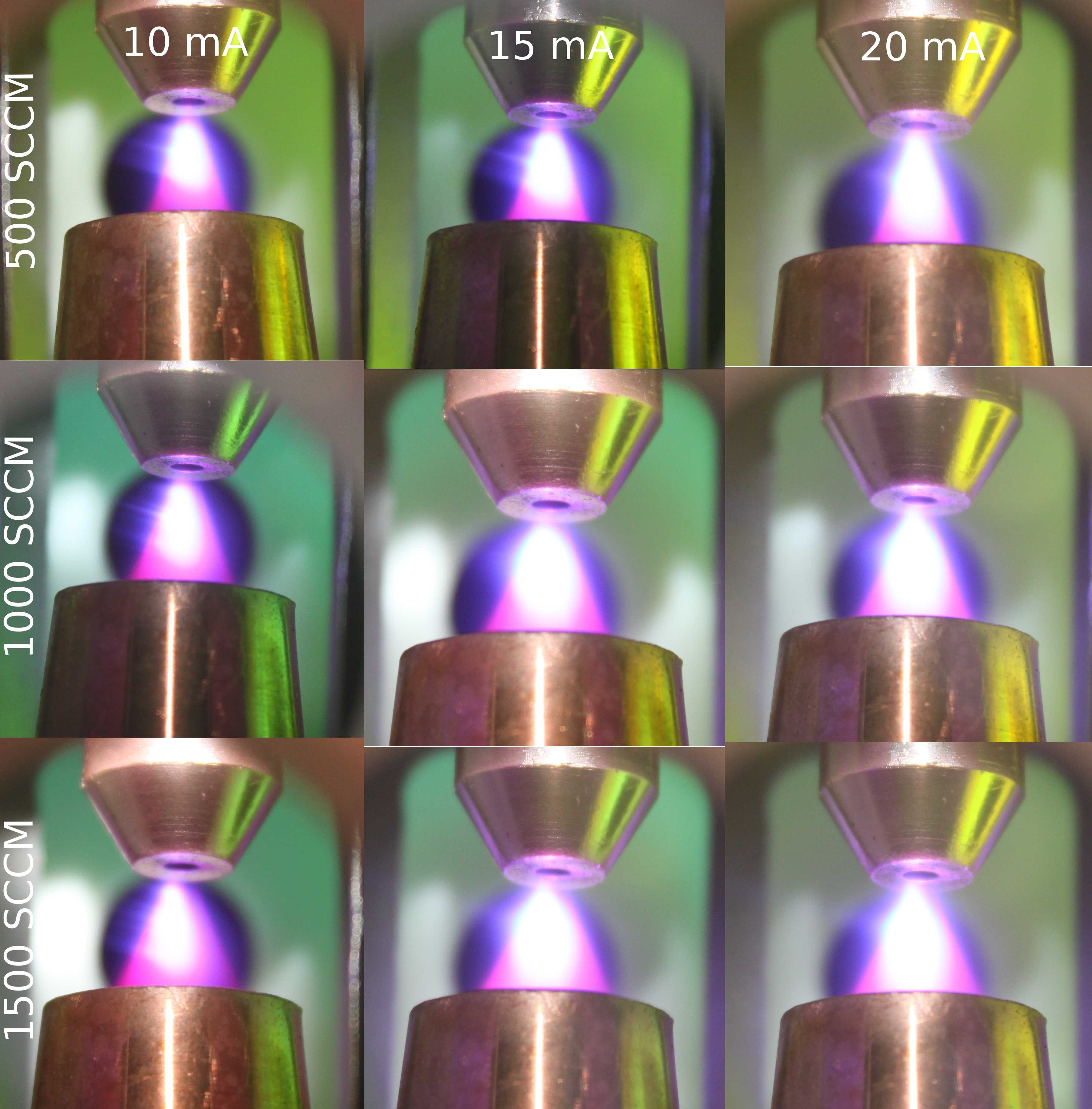}
  \end{center}
  \caption{Pictures of the VUV-spot with different settings of the gas flow rate (rows: $500$, $1000$ and $1500$ sccm) and the emission current (columns: $10$, $15$ and $20$ mA).}
  \label{spot_labels}
\end{figure*}

Fig. \ref{spot_labels} illustrates the size and intensity of the spot qualitatively with different settings for the emission current and for the gas flow. The photos were taken through the window at the opposite side of the light outlet (see Fig. \ref{vuv_2}).

Based on the experience of the first design of the VUV simulator \citep{ver}, the same gas mixture has been chosen to produce the spectra. The mixture provides a continuum spectrum similar to the Sun's spectrum. Due to the presence of the water particles in the chamber the Hydrogen Lyman $\alpha$ as well as the other H lines are present (see Figs. \ref{s1} and \ref{s2}). The intensity of the VUV light increases generally up to a saturation value with increasing gas flow. An increasing gas flow, however, reduces the quality of the vacuum in the radiation chamber. Thus, the sensitive balance between VUV light intensity and vacuum quality has to be taken into account when optimizing the VUV source parameters. In the same way, a larger emission current causes, for a given gas flow, a higher intensity of the radiation. Therefore it is favorable to operate the simulator with relatively small gas flows and high electron currents to get the same intensity.

After first function tests concerning the stability and operating life, the following settings were chosen for the calibration procedure:

\begin{itemize}
  \item Gas mixture: Ar ($98.5$\%), Kr ($1$\%) and He ($0.5$\%)
  \item Electron energy: $1$ keV,
  \item Electron emission current: $20$ mA,
  \item Flow rate of the gas jet: $300$, $600$, $1200$ sccm.
\end{itemize}

The goals for the calibration procedure were:

\begin{enumerate}
  \item Finding the optimal alignment between electron beam and gas jet,
  \item Finding the magnetic lens current for an optimal focus of the electron beam,
  \item Measurement of the spectral radiant intensity of the radiation.
\end{enumerate}

\section{The method and procedure of calibration}
\label{calibration}

The VUV source was calibrated in the radiometry laboratory of the Physikalisch-Technische Bundesanstalt (PTB) at Berlin Electron Storage Ring for Synchrotron Radiation of Helmholtz Zentrum Berlin (BESSY II) in Berlin-Adlershof. PTB provides the calibration of radiation detectors and sources as well as the characterization of optical components in the UV and VUV range \citep{ptb1,ptb2}. 

\begin{figure*}[!t]
  \begin{center}
    \includegraphics*[width=0.7\textwidth]{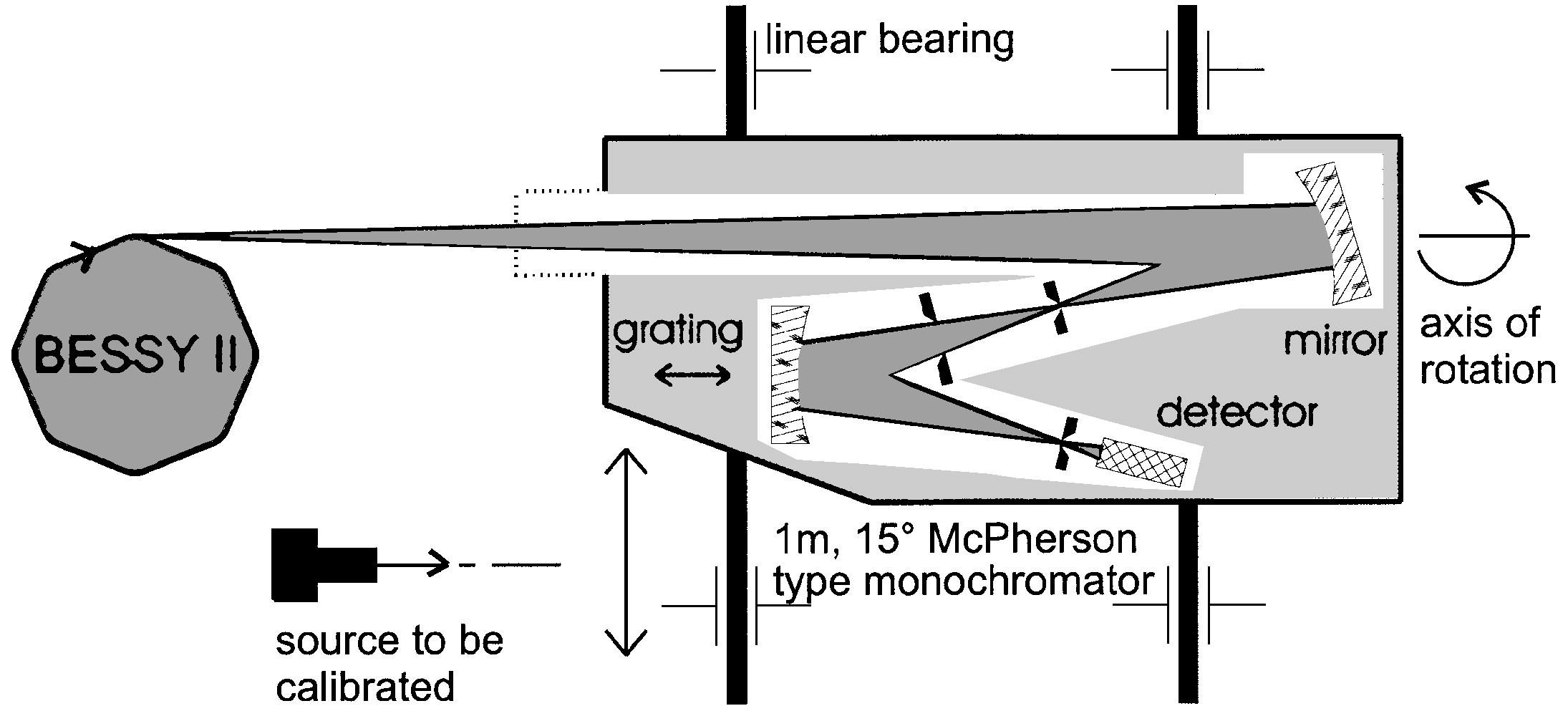}
  \end{center}
  \caption{The normal - incidence monochromator beam line used for calibration of radiation sources \citep{ptb1}.}
  \label{ptb_calibration}
\end{figure*}

Fig. \ref{ptb_calibration} shows the normal-incidence monochromator  ($1$m, $15^{\rm{o}}$ McPherson type) beam line for source calibration in the configuration for the calibration of the measurement site, where synchrotron radiation is used as a primary source standard.

The whole device can be rotated around the entrance axis to take the polarization of synchrotron radiation into account during that procedure. Moreover, it is possible to shift the monochromator compartment via linear bearings perpendicular to the entrance axis in order to connect it with the light outlet of the source to be calibrated. 

The toroidal mirror (see Fig. \ref{ptb_calibration}) images the light spot into the entrance slit of the monochromator. The solid angle of accepted radiation is precisely defined by apertures (not shown in Fig. \ref{ptb_calibration}). A spherical grating maps the entrance slit in the first spectral order into the exit slit, where the radiation of the selected wavelength is detected by a photomultiplier tube. The corresponding wavelength and the spectral resolution depend on the angular position of the grating, the grating constant, the slit width, and other parameters, which are optimized in order to suppress the influence of higher spectral order too. The measurement of the spectral radiant intensity is performed by varying the angle of the grating in different spectral ranges with a given resolution (wavelength scan). Additionally, it is possible to vary the angle of the toroidal mirror to measure the intensity at different horizontal and vertical positions at a fixed wavelength (in combination with appropriate apertures and slits) to record a spatial profile of the radiating spot and to align the image of the spot into the entrance slit of the monochromator. Four monochromator configurations were used to calibrate the VUV-simulator in the following spectral ranges with given resolutions:

\begin{itemize}
  \item $40$ nm to $120$ nm, resolution: $0.2$ nm,
  \item $110$ nm to $220$ nm, resolution: $0.2$ nm,
  \item $160$ nm to $330$ nm, resolution: $1$ nm,
  \item $300$ nm to $410$ nm, resolution: $1$ nm.
\end{itemize}

The first step for the calibration of the VUV-simulator was the alignment to the entrance axis of the monochromator and the connection of the vacuum systems of both facilities. After first wavelength scans in the range between $110$ nm and $220$ nm a relatively intense spectral line was chosen for the alignment of the electron beam to the gas jet. That means, the monochromator was adjusted to that intense wavelength (e.g. $123.62$ nm) and the orientation of the electron source and the gas nozzle were changed during the timely monitoring of the detector signal to find the optimum. The same procedure was performed for the setting of the magnetic lens current. After that, the wavelength scans were performed with the settings of the VUV-simulator described in Section \ref{vuv} in the spectral ranges given above.

\section{The VUV spectra, comparison to the solar spectra}   

\subsection{Spectral intensity distribution}
\label{spectra}

In order to get the spectral radiant intensity in W $\rm{sr^{-1}}$ $\rm{nm^{-1}}$, the measured detector signals in the described wavelength ranges were converted with the corresponding measurement site sensivity by PTB. By use of the geometrical parameters of the CIF that spectral radiant intensity was re-calculated into spectral irradiance at the position of the object under test in W $\rm{m^{-2}}$ $\rm{nm^{-1}}$. Since for each parameter setting several measurements have been performed, an averaged spectral distribution was calculated (see Section \ref{stability}).

\begin{figure*}[!t]
  \begin{center}
    \includegraphics*[angle=90, width=0.6\textwidth]{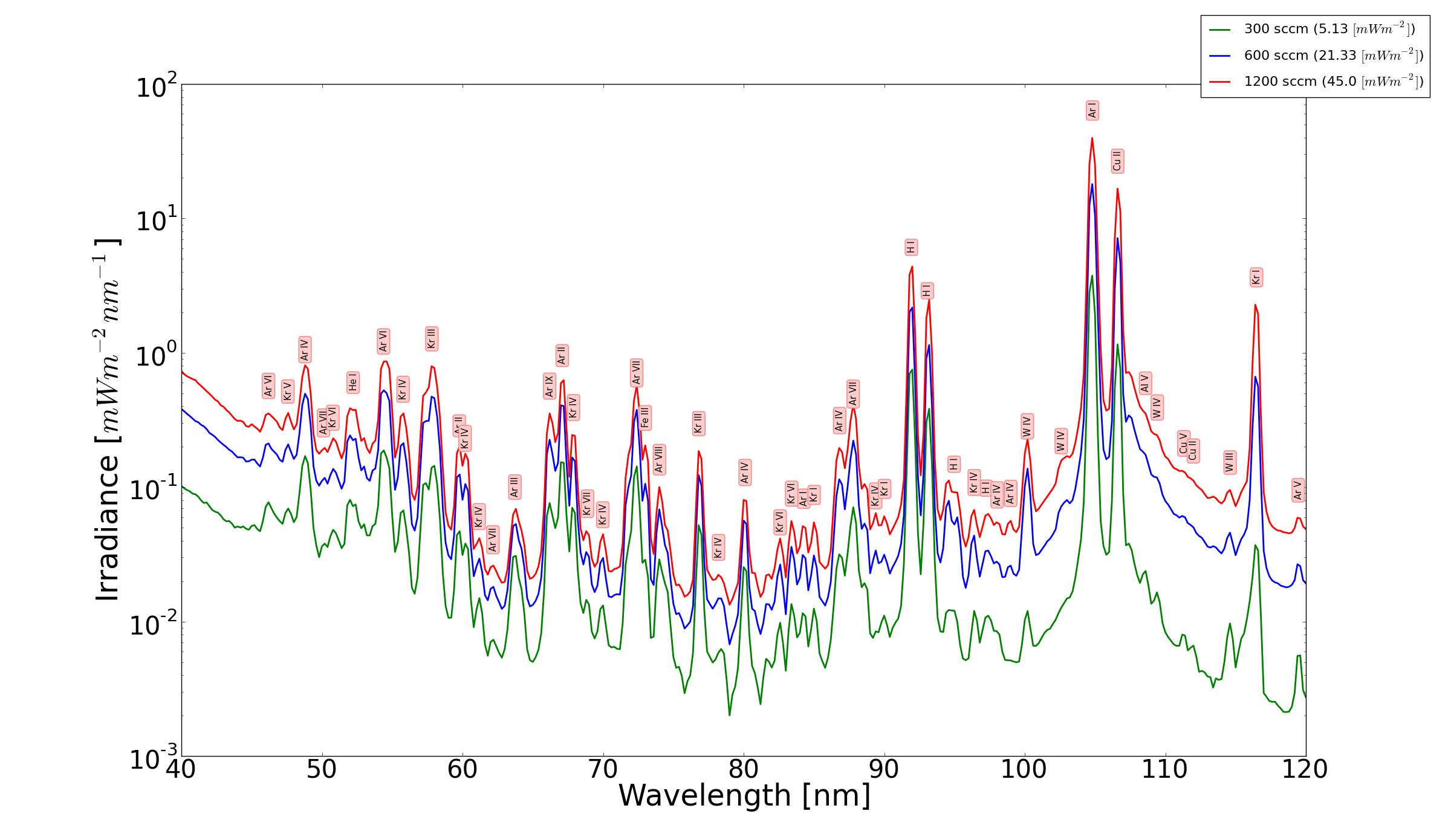}
  \end{center}
  \caption{The VUV spectral lines in the wavelength range from $40$ nm to $120$ nm for different gas flows: $300$ (green line), $600$ (blue line) and $1200$ (red line).}
  \label{s1}
\end{figure*}

\begin{figure*}[!t]
  \begin{center}
    \includegraphics*[angle=90, width=0.6\textwidth]{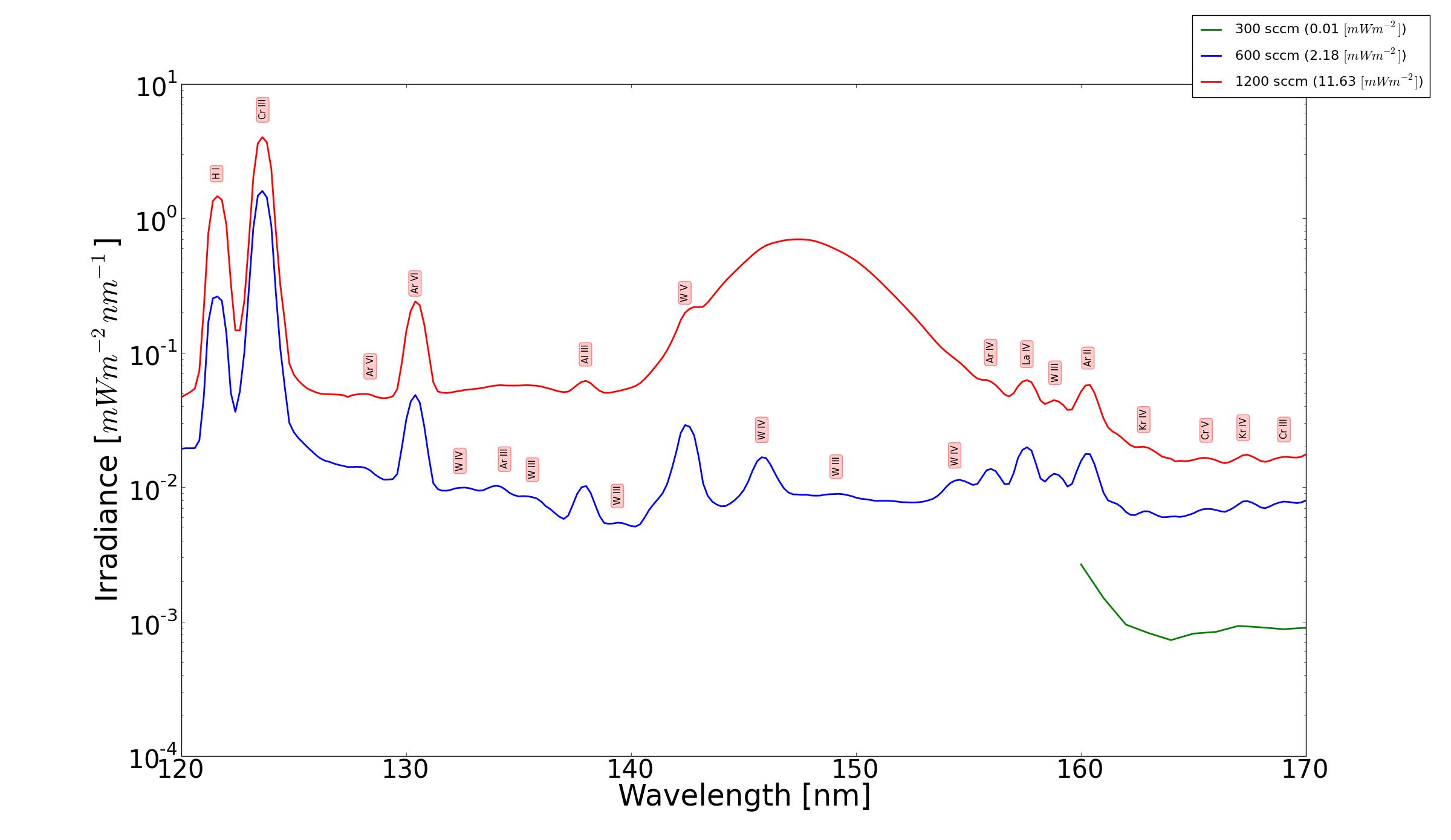}
  \end{center}
  \caption{The VUV spectral lines in the wavelength range from $120$ nm to $170$ nm for different gas flows: $300$ (green line), $600$ (blue line), $1200$ (red line).}
  \label{s2}
\end{figure*}

\begin{figure*}[!t]
  \begin{center}
    \includegraphics*[angle=90, width=0.6\textwidth]{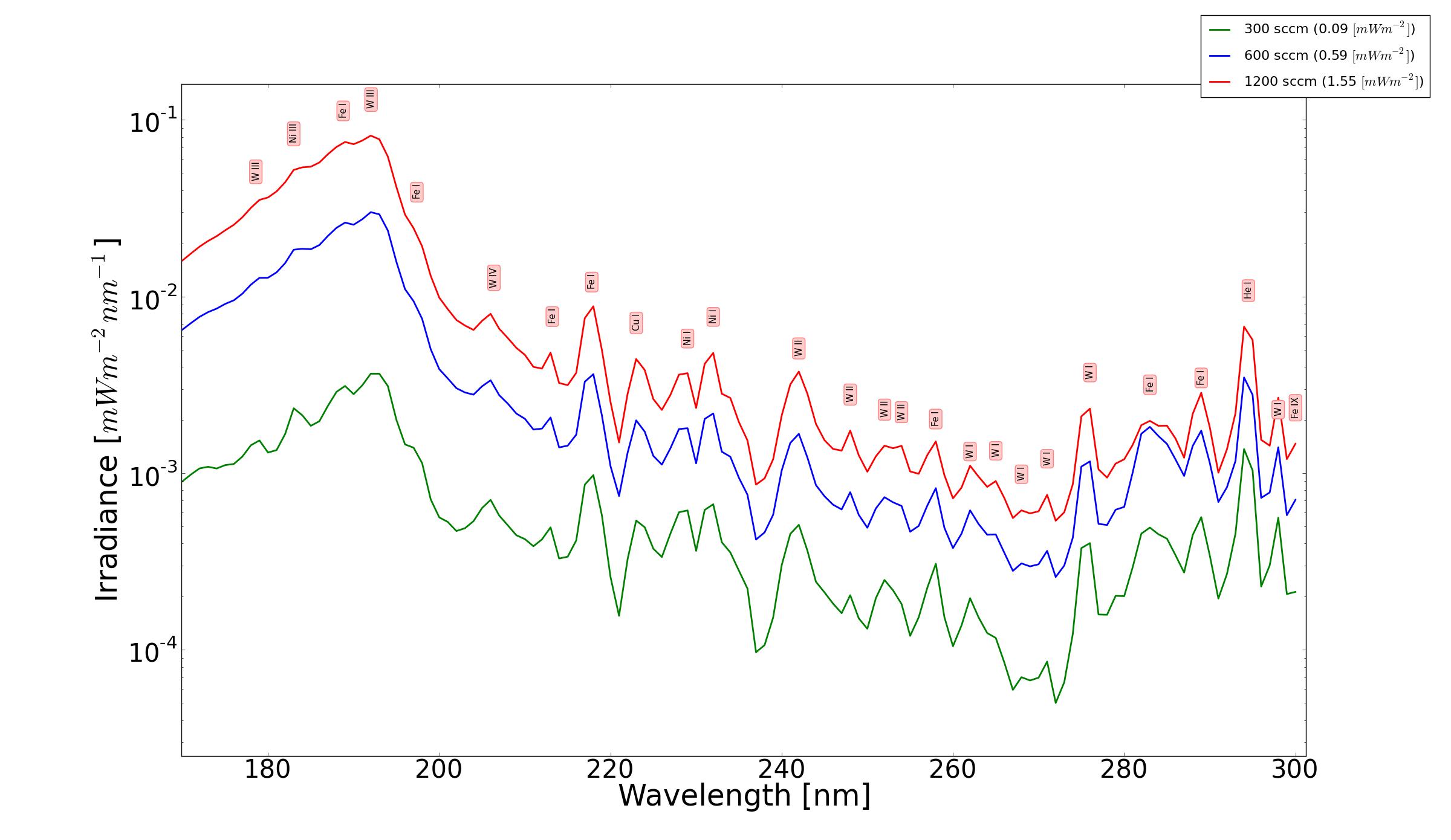}
  \end{center}
  \caption{The VUV spectral lines in the wavelength range from $170$ nm to $300$ nm for different gas flows: $300$ (green line), $600$ (blue line), $1200$ (red line).}
  \label{s3}
\end{figure*}

\begin{figure*}[!t]
  \begin{center}
    \includegraphics*[angle=90, width=0.6\textwidth]{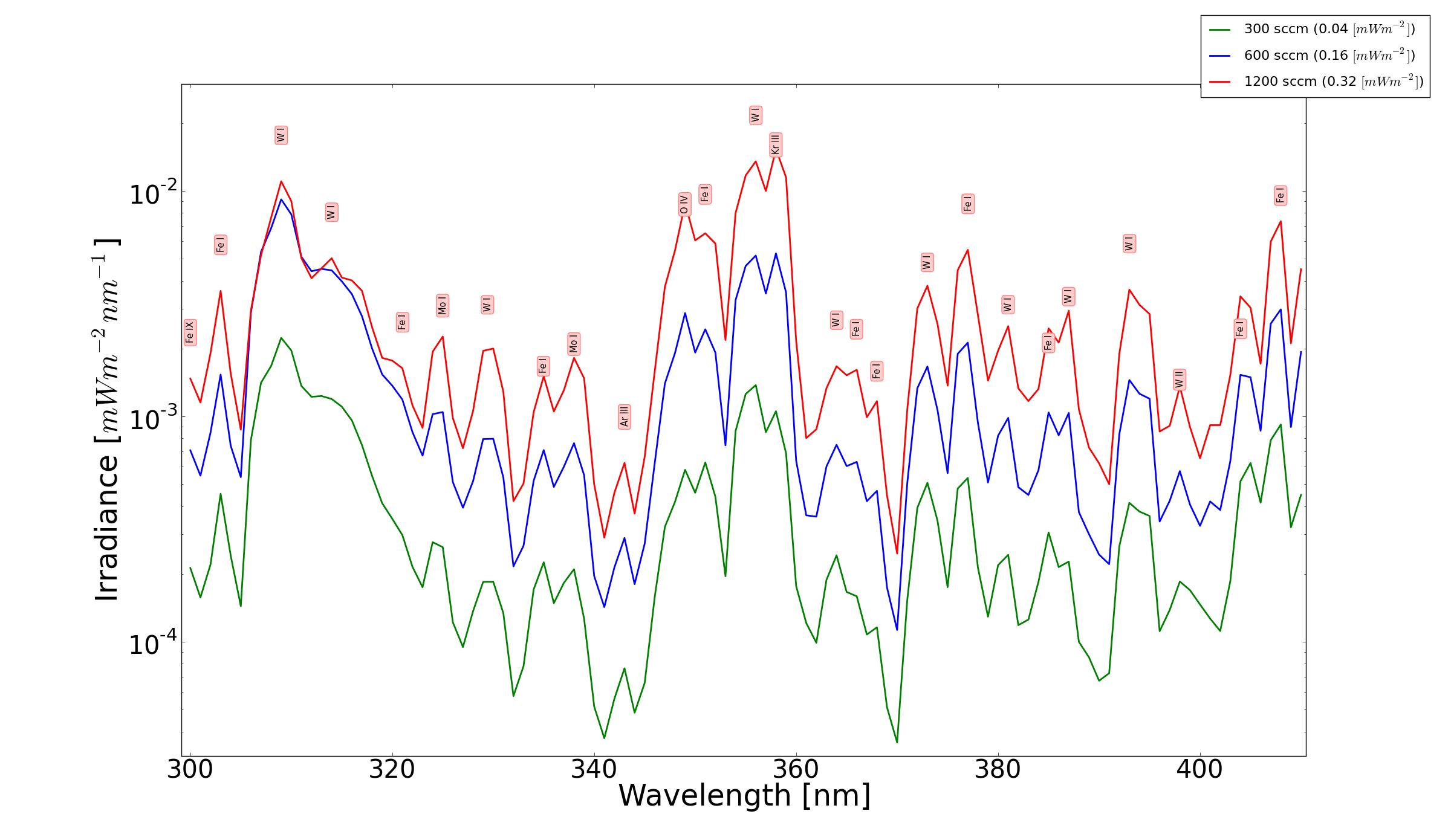}
  \end{center}
  \caption{The VUV spectral lines in the wavelength range from $300$ nm to $410$ nm for different gas flows: $300$ (green line), $600$ (blue line), $1200$ (red line).}
  \label{s4}
\end{figure*}

\begin{figure*}[!t]
  \begin{center}
    \includegraphics*[angle=90, width=0.6\textwidth]{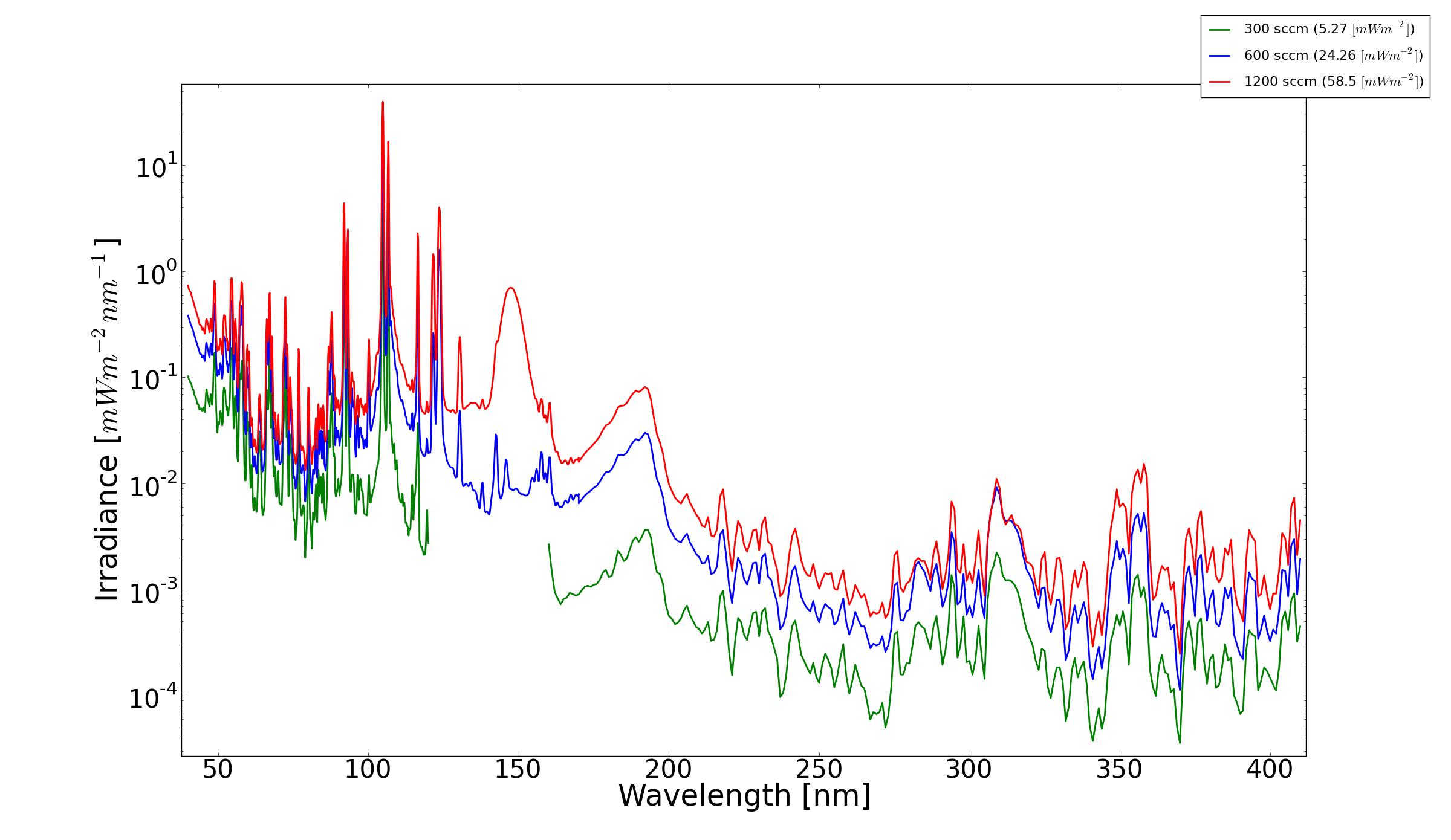}
  \end{center}
  \caption{Summary plot of the VUV spectra in the wavelength range from $40$ nm to $410$ nm for different gas flows: $300$ (green line), $600$ (blue line), $1200$ sccm (red line).}
  \label{s_summary}
\end{figure*}

The resulting spectral irradiance distributions are shown in Figs. \ref{s1}-\ref{s4} for each gas flow ($300$, $600$, $1200$ sccm) with a different colour. The integral irradiance for each configuration of the VUV-source is given in the legend. A survey of all spectra over the whole range from $40$ to $410$ nm is presented in Fig. \ref{s_summary}.

The spectra of the VUV-simulator are characterized by a large number of lines. The verification of these spectral lines was made by use of the database of the National Institute of Standards and Technology (NIST) \citep{nist}.
Each identified spectral line is marked in Figs. \ref{s1}-\ref{s4} by a label which contains the name of the chemical element and the degree of ionization. 
An overview of all identified spectral lines is presented in the Appendix (Table \ref{tab:1}).

Spectral lines in the wavelength range from $138$ nm to $160$ nm  are classified separately (see Fig. \ref{s2} and Table \ref{tab:2}). For larger gas flows the spectral lines in that range disappear and large bumps appear. This is an effect of the increasing number of collisions between the gas atoms. That result in a broadening of the spectral lines which form eventually a continuum. The NIST database possesses in this wavelength range $112$ Ar lines, $140$ Kr lines and no He lines.

In order to get an idea how good the VUV-source simulates the solar spectrum, the appearance of solar spectral lines and their possible counterparts of the VUV-simulator has to be studied.

\begin{figure*}[!t]
  \begin{center}
    \includegraphics*[angle=90, width=0.6\textwidth]{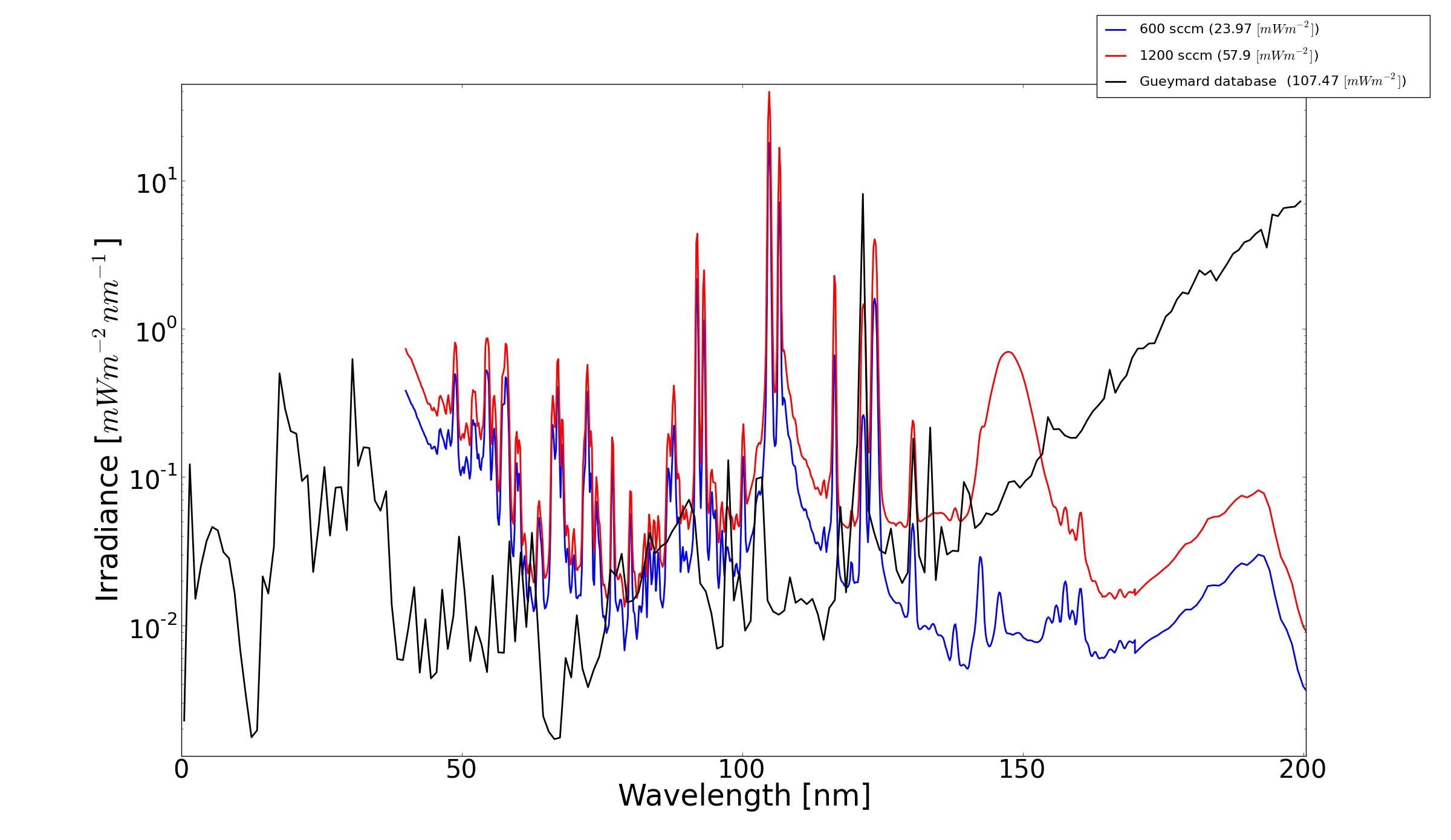}
  \end{center}
  \caption{The spectra of the VUV-simulator for different gas flows ($600$ and $1200$ sccm) comparing to the Gueymard database denoted as a black line. The given total intensity of the solar spectrum taken from the Gueymard database is calculated from $40$ nm to $200$ nm and it is shown in the legend.}
  \label{e490_vuv}
\end{figure*}

Since the ASTM E-490 standard of the spectral solar emissivity covers the wavelength range from $119.5$ nm to $1000000$ nm, the Gueymard database \citep{guey} is used as a reference for wavelengths $< 120$nm. The VUV-simulator spectrum and the Gueymard standard are shown in Fig. \ref{e490_vuv}. From $119.5$ nm to $150$ nm the VUV-simulator spectrum coincides well with the solar emission lines. Below $119.5$ nm the spectral intensity of the VUV-simulator is larger than the solar one. Since the VUV-source has a very small intensity for wavelengths above $150$ nm, this range must be covered by the light of a Deuterium lamp - another light source of the CIF.

To compare the VUV-simulator emission lines with the solar ones, the SUMER database \citep{sumer} has been chosen as a reference. A Lyman-$\rm{\alpha}$ line at $121.57$ nm has been identified both in the spectrum of the VUV-simulator and in \citep{sumer}. Its intensity in the spectrum of the VUV-simulator is $18$\% of the corresponding solar one. Strong Hydrogen lines of the VUV source are present at: $91.93$ nm ($131$ times stronger-), $93.07$ nm ($138$ times stronger-), and $94.97$ nm ($10$ times stronger- than the Sun spectrum at that wavelengths). The presence of Hydrogen and residual water vapor in the VUV-simulator is a consequence of the fact, as it is the predominant residual gas in metal vacuum systems at very low pressures \citep{redhead}. The two strongest lines of the VUV-source spectrum appear at $104.82$ nm and $106.61$ nm. They correspond to the Ar I and Cu II lines that are $2922$ - times and $1423$ - times stronger than the solar spectrum at these wavelengths. In the solar spectrum these lines are not present. Also not present in the solar spectrum are the VUV-source lines at $123.62$ nm and $116.49$ nm (Kr I and Cr III). The Ar- Kr- and He-lines are caused by constituents of the gas flow excited by the electron beam. The metal lines such as: Cu, Cr, W, Fe, Mo, Ni, La and Al originate from the construction materials of the VUV-simulator. These lines are constituents of the solar spectrum too \citep{sumer}. 

\begin{figure*}[!t]
  \begin{center}
    \includegraphics*[angle=90, width=0.6\textwidth]{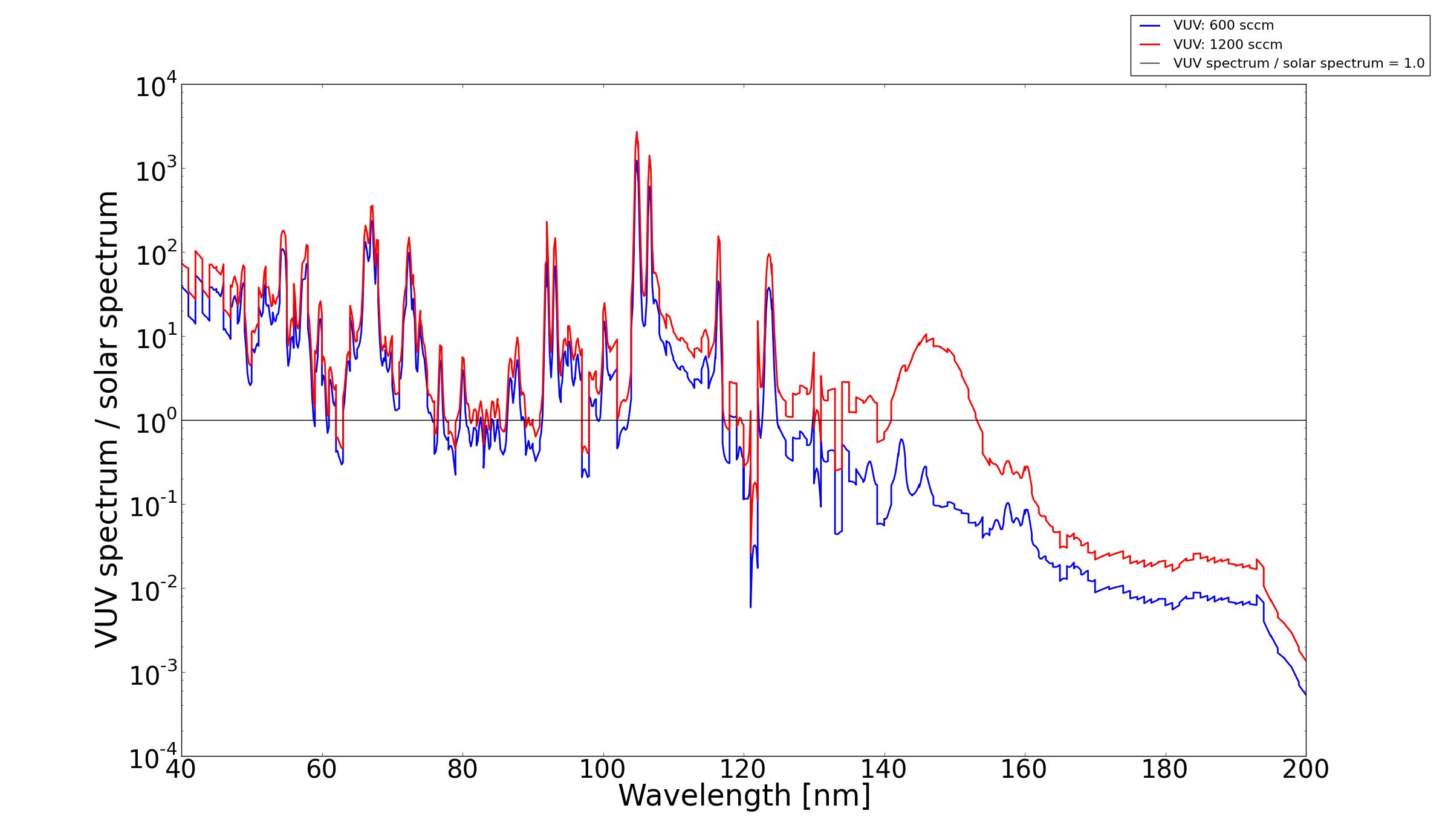}
  \end{center}
  \caption{VUV spectra for gas flows of $600$ and $1200$ sccm divided by the spectrum of the Sun taken from the Gueymard database \citep{guey}.}
  \label{sun_sources}
\end{figure*}

\subsection{Acceleration factors}
\label{acc}

The acceleration factor of material tests for space application is generally defined as the ratio of the intensity of a degrading radiation applied to a material at the laboratory versus the intensity of the same degrading factor in space \citep{ecss}. In the following the acceleration factors achieved by the VUV-simulator are discussed. They are calculated as the ratio of its intensity in a certain spectral range to the intensity of the solar radiation in the same range at $1$ AU. The factors decrease if materials are tested for space applications that go closer to the Sun, whereas the factors increase in the case of applications that veer away from the Sun. 

The differential as well as integrated values of the Gueymard's and VUV-simulator's spectral intensities and acceleration factors are presented in Tables \ref{tab:short} and \ref{tab:long}, respectively. The highest differential acceleration factor of $95.4$ SC is reached in the wavelength range $100$ nm to $110$ nm and a gas flow of $1200$ sccm (see Table \ref{tab:short}). The integral of the VUV-simulator's irradiance can reach: $26.3$ SC for a gas flow of $1200$ sccm, $12.5$ SC for a gas flow of $600$ sccm and $3$ SC for a gas flow of $300$ sccm in the wavelength range from $40$ nm to $120$ nm. Fig. \ref{e490_vuv} shows the spectra of the VUV-simulator for two different gas flows $600$ and $1200$ sccm. There is plotted the solar spectrum of the Gueymard database for comparison too. The solar spectral intensity is larger than that of the VUV-simulator at wavelengths higher than $150$ nm.

The spectral distribution of the accelerating factors is presented in Fig. \ref{sun_sources}. The spectral intensity distributions of the VUV-simulator are divided by the solar ones taken from the Gueymard database for two representative gas flows. The acceleration factor is significantly larger than 1 in almost the complete range up to $120$ nm. For wavelengths higher than $115$ nm, a Deuterium lamp yields higher intensities and sufficiently large acceleration factors. Depending on the specific purposes of experiments to determine degradation effects of materials exposed to VUV radiation, the appropriate gas flow has to be chosen.

\begin{table*}[!ht]
\centering
\caption{The differential values of the Gueymard's and VUV-simulator's spectral intensities as well as acceleration factors in bins of $10$ nm. Unfortunately, there is no matchable spectrum for a gas flow of $300$ sccm in the wavelength range of $120-160$ nm.}
\label{tab:short}
\scalebox{0.75}{\begin{tabular}{l|l|l|l|l|l|l|l} 
  \noalign{\smallskip}\hline\noalign{\smallskip}
   Wavelength & Gueymard   & \multicolumn{2}{c}{300 sccm} & \multicolumn{2}{|c}{600 sccm} & \multicolumn{2}{|c}{1200 sccm}\\
  nm & mW $\rm{m^{-2}}$ & mW $\rm{m^{-2}}$ & Acc. & mW $\rm{m^{-2}}$ & Acc. & mW $\rm{m^{-2}}$ & Acc. \\
\noalign{\smallskip}\hline\noalign{\smallskip}
  40-50  & 0.11 & 0.69  & 6.27  & 2.23  & 20.27  & 4.05  &  36.82\\
  50-60  & 0.14 & 0.62  & 4.43  & 1.85 & 13.21  & 3.09  &  22.07\\
  60-70  & 0.11 & 0.28 & 2.55 & 0.69  & 6.27  & 1.08  &  9.82\\
  70-80  & 0.13 & 0.19 & 1.46 & 0.49  & 3.77  & 0.79  &  6.08 \\
  80-90  & 0.33 & 0.14 & 0.42  & 0.42  & 1.27  & 0.76  & 2.30  \\
  90-100 & 0.37 & 0.63 & 1.70 & 1.85  & 5.00  & 3.74  & 10.11 \\
  100-110& 0.31 & 2.51 & 8.10  & 13.11 & 42.29 & 29.56 & 95.35 \\
  110-120& 0.21 & 0.07 & 0.33  & 0.07  & 3.33  & 1.93 & 9.19 \\
  120-130& 8.58 & -    & -     & 1.77 & 0.21  & 5.24 & 0.61 \\
  130-140& 0.67 & -    & -     & 0.12 & 0.18  & 0.70 & 1.04 \\
  140-150& 0.69 & -    & -     & 0.11 & 0.16  & 4.02 & 5.83 \\
  150-160& 1.65 & -    & -     & 0.11 & 0.07  & 1.46 & 0.88 \\
  160-170& 3.60 & 0.01 & $2.78\times10^{-3}$ & 0.08 & 0.02  & 0.22 & 0.06 \\
  170-180& 11.09& 0.01 & $9.02\times10^{-4}$ & 0.08 & $7.21\times10^{-3}$ & 0.21 & 0.02\\
  180-190& 26.01& 0.02 & $7.69\times10^{-4}$ & 0.18 & $6.92\times10^{-3}$ & 0.53 & 0.02\\
  190-200& 53.47& 0.02 & $3.74\times10^{-4}$ & 0.19 & $3.55\times10^{-3}$ & 0.53 & 0.01\\
\noalign{\smallskip}\hline\noalign{\smallskip}
\end{tabular}}
\end{table*}

\begin{table*}[!ht]
\centering
\caption{The integrated values of the Gueymard's and VUV-simulator's spectral intensities as well as acceleration factors with fixed lower limit of the wavelength range ($40$ nm). Unfortunately, there is no matchable spectrum for a gas flow of $300$ sccm in the wavelength range of $120-160$ nm. }
\label{tab:long}
\scalebox{0.8}{\begin{tabular}{l|l|l|l|l|l|l|l} 
  \noalign{\smallskip}\hline\noalign{\smallskip}
   Wavelength & Gueymard   & \multicolumn{2}{c}{300 sccm} & \multicolumn{2}{|c}{600 sccm} & \multicolumn{2}{|c}{1200 sccm}\\
  nm & mW $\rm{m^{-2}}$ & mW $\rm{m^{-2}}$ & Acc. & mW $\rm{m^{-2}}$ & Acc. & mW $\rm{m^{-2}}$ & Acc. \\
\noalign{\smallskip}\hline\noalign{\smallskip}
  40-50  & 0.11 & 0.69 & 6.27  & 2.23 & 20.27 & 4.05 & 36.82  \\
  40-60  & 0.25 & 1.31 & 5.24 & 4.08 & 16.32 & 7.14 & 28.56 \\
  40-70  & 0.36 & 1.59 & 4.42 & 4.77 & 13.25 & 8.22 & 22.83 \\
  40-80  & 0.49 & 1.78 & 3.63 & 5.26 & 10.73 & 9.01 & 18.39 \\
  40-90  & 0.82 & 1.92 & 2.34 & 5.68 & 6.93 & 9.77 & 11.91 \\
  40-100 & 1.19 & 2.55 & 2.14  & 7.52 & 6.32 & 13.51& 11.35  \\
  40-110 & 1.50 & 5.06 & 3.37 & 20.64& 13.76 & 43.07& 28.71 \\
  40-120 & 1.71 & 5.13 & 3.00 & 21.33& 12.47  & 45.00& 26.32 \\
  40-130 & 10.29& - & -& 23.11& 2.25 & 50.25 & 4.88 \\ 
  40-140 & 10.96& - & -& 23.22& 2.12 & 50.95 & 4.65 \\
  40-150 & 11.65& - & -& 23.33& 2.00 & 54.97 & 4.72 \\
  40-160 & 13.30& - & -& 23.44& 1.76 & 56.43 & 4.24 \\
  40-170 & 16.90& - & -& 23.52& 1.39 & 56.64 & 3.35 \\
  40-180 & 27.99 & -& -& 23.60& 0.84 & 56.86 & 2.03 \\
  40-190 & 54.00 & -& -& 23.78& 0.44 & 57.39 & 1.06 \\
  40-200 & 107.47& -& -& 23.98& 0.22 & 57.92 & 0.54 \\
\noalign{\smallskip}\hline\noalign{\smallskip}
\end{tabular}}
\end{table*}

\subsection{The stability of the radiation intensity of the VUV-simulator}
\label{stability}

The presumption of the variability of the radiation intensity, as seen during the calibration campaign, has been confirmed by the stability analysis. In order to get an idea about the stability, several measurements with identical settings of gas flow and electron current has been performed at different days and different operating times after the regeneration of the source during the calibration process at PTB.

The stability is affected by different effects which may compensate each other, at least partially. One effect is the formation of ice at the cold baffles around the light spot. The growing lumps of ice decrease the pumping power. Therefore, the intensity could increase because the gas density increases. It has to be mentioned, that this increase of intensity is mostly seen in measurements performed at the same day, however, not always and not in the full spectral range. On the other hand, the apertures at the cold baffles for the electron beam and for the light outlet will freeze up at longer operating time. That may cause a decrease of intensity. It will be reduced by the lower and/or defocussed electron beam and/or the lower solid angle of the out-coming light. Other the stability influencing effects are a small variation of both the gas flow and the electron current.

Due to the small number of comparable measurements a complete statistical analysis was not feasible. Instead, a worst case estimation has been performed as follows.

\begin{figure*}[!t]
  \begin{center}
    \includegraphics*[width=0.8\textwidth]{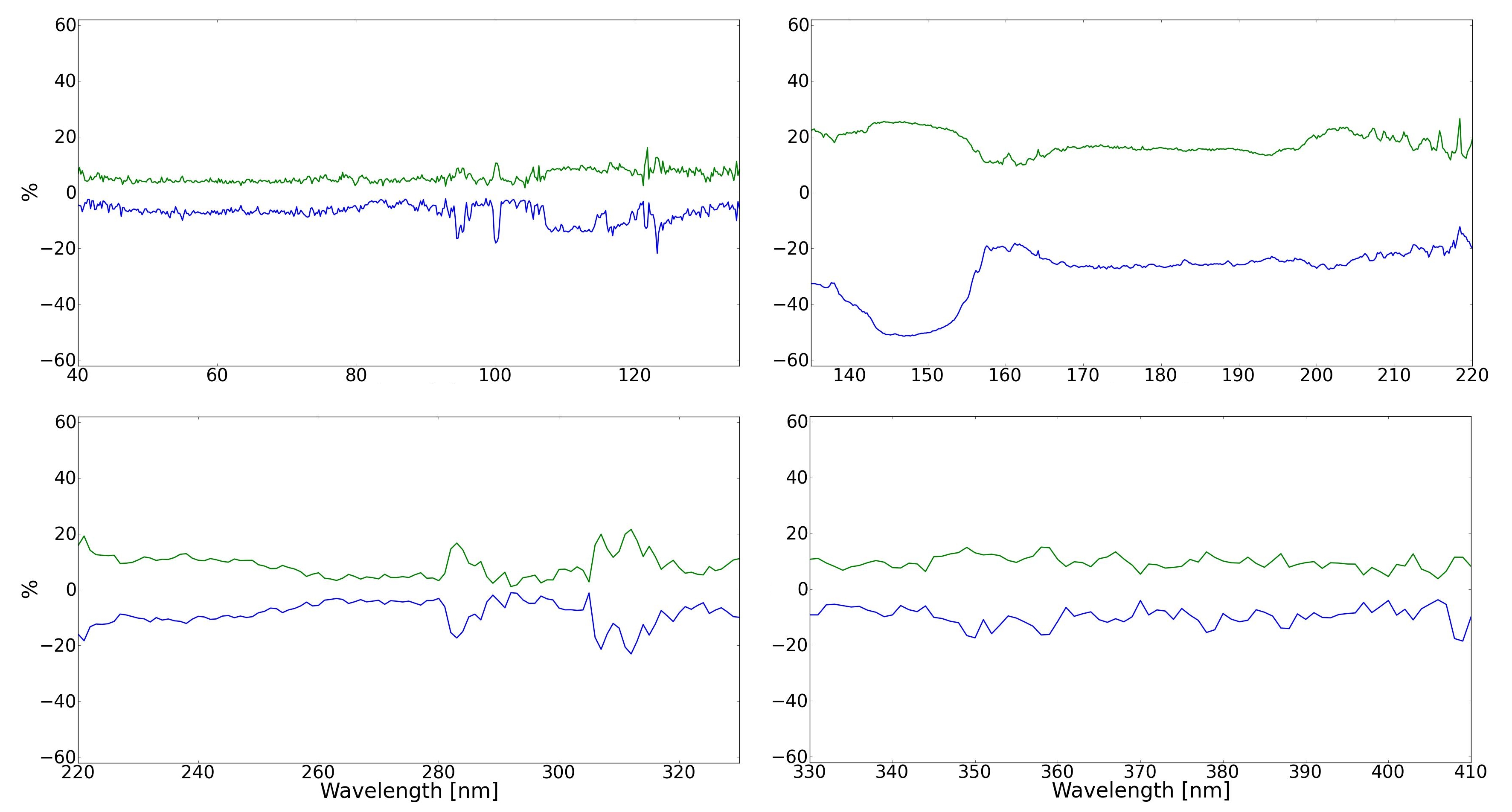}
  \end{center}
  \caption{The maximum deviation (green line positive, blue line negative) of the spectral intensity distribution in the four spectral ranges related to their mean value.}
  \label{summary_final}
\end{figure*}

The stability of the VUV-simulator has been estimated by comparing the intensity spectra taken at almost identical settings but made at as far as possible distant moments in time. The different number of measurements at each spectral range ($40-135$ nm: $3$; $135-220$ nm: $9$; $220-330$ nm: $5$; $330-410$ nm: $5$) was taken into account to determine the mean value. The maximum deviation was calculated by division of the maximum respectively the minimum intensity by the mean value at each wavelength. The results are shown in Fig. \ref{summary_final} only for the gas flow of 1200 sccm. 

The lowest deviation to the average signal of +/- 10\% appears in the wavelength range from $40$ nm to $105$ nm. Above $105$ nm to $135$ nm are bands where the deviation is +/- $20$\% (see upper left plot). The highest one of $-50$\% is in the wavelength range of $145$ nm to $155$ nm, while in the range of $160$ nm to $220$ nm the deviation is +/-$25$\% (see upper right plot). In the ranges of $220$ nm to $330$ nm and $330$ nm to $410$ nm the deviation from the average signal is equal or less than +/- $20$\% (see lower left and right plot).

The analysis of the plots in Fig. \ref{summary_final} shows an increase of the maximum deviation in the higher wavelength ranges. This result is influenced by the different number of measurements too. Nevertheless, it is acceptable in the range of short wavelengths. This stability analysis shows that the VUV simulator is not qualified as a calibration standard, i.e. for detector calibration. However, the stability in the VUV range is sufficient to perform irradiation tests for material investigations. The larger deviations in higher wavelength ranges are not a serious problem since the intensity there is at least one order of magnitude smaller than in the range below $115$ nm see Figs \ref{s_summary} and \ref{e490_vuv}. The very low intensity of the VUV-simulator at wavelengths larger than 115 nm is in the CIF compensated by the Deuterium lamp. It exceeds for wavelengths $>115$ nm the solar intensity by about one order of magnitude.

\section{Summary}
\label{summary}
Short wavelength electromagnetic radiation as generated by the VUV-simulator plays a crucial role in space material science due to photoionization and photodissociation effects (see Section \ref{intro}). The maximum acceleration factor reached at the wavelength range from $40$ nm to $120$ nm is about $26.3$ SC at a gas flow of $1200$ sccm. In the same range but at a gas flow of $600$ sccm the acceleration factor is $12.5$ SC. With the smallest gas flow used in the calibration procedure ($300$ sccm) the factor is about $3$ SC. Since the source has many operational parameters (see Section \ref{vuv}), changes of the gas flow will cause variations of the acceleration factor. Given an operating time of at least 8 h and the large acceleration factors, the VUV-simulator is a suitable facility for various material tests and degradation experiments.    

The simulator passed the first campaign of the calibration procedure. The spectral lines are calibrated from $40$ nm to $410$ nm. Based on the experiences made with first VUV-simulator \citep{ver} a significant intensity down to $5$ nm can be expected, i.e. this source would cover also the soft X-ray range of the solar spectrum. Therefore a second calibration campaign for wavelengths smaller than $40$ nm is necessary. It can be performed at the recently set-up facility for source calibration at the Metrology Light Source of PTB (MLS) \citep{gott}.

The calibration procedure has proved that the VUV-simulator meets the requirements with respect to the solar spectral intensity distribution, the achievable acceleration factors, and the size of the irradiated area.

The stability analysis of the VUV-simulator signal shows that the maximum intensity deviations in the VUV range below $115$ nm are in the order of $10$\%. The larger deviation for wavelengths above 115 nm are not a serious problem for material testing in the CIF because the VUV-simulator intensity in that range is negligible small and the Deuterium lamp is used (see Section \ref{stability}). Therefore, a satisfactory operation of the VUV-simulator can be expected. Nevertheless, in a forthcoming calibration campaign the stability will be subject of a more systematic analysis.

\section{Acknowledgements}
We want to acknowledge the big help of Simone Kroth and Wolfgang Paustian at their assistance in performing all the measurements for conditioning and calibrating the source.

\appendix
\section{Tabulated spectra lines}

Table \ref{tab:1} contains validated spectra lines and their elements in the wavelength range from $40$ to $410$ nm. The continuous spectrum in the wavelength range from $138$ to $160$ nm is presented in the Table \ref{tab:2}. All of the spectra lines and their elements are taken from the NIST database \citep{nist}.

\newpage
\begin{table*}[!ht]
\centering
\caption{Validated spectra lines in the wavelength range from $40$ to $410$ nm. The lines are depicted in Figs. \ref{s1}, \ref{s2}, \ref{s3} and \ref{s4}. Data are taken from the NIST database \citep{nist}.}
\label{tab:1}
\scalebox{1.0}{\begin{tabular}{ll|ll|ll} 
  \noalign{\smallskip}\hline\noalign{\smallskip}
Wavelength [nm] & Element & Wavelength [nm] & Element & Wavelength [nm] & Element   \\
\noalign{\smallskip}\hline\noalign{\smallskip}
46.201  & Ar VI  & 104.822   & Ar I   &  248.00368 & W II   \\
47.575  & Kr V   & 106.6134  & Cu II  &  252.00734 & W II   \\
48.799  & Ar IV  & 108.592   & Al V   &  253.99098 & W II   \\
50.109  & Ar VII & 109.4264  & W IV   &  257.98438 & Fe I   \\
50.782  & Kr VI  & 111.322   & Cu V   &	 261.9954  & W I    \\
52.2186 & He I   & 111.9947  & Cu II  &	 264.998   & W I    \\
54.383  & Ar VI  & 114.6102  & W III  &	 268.0046  & W I    \\
55.775  & Kr IV  & 116.4867  & Kr I   &	 271.0004  & W I    \\
57.8212 & Kr III & 119.4528  & Ar V   &	 276.0036  & W I    \\
59.7701 & Ar II  & 121.56699 & HI     &	 283.0054  & Fe I   \\
60.183  & Kr IV  & 123.62    & Cr III &	 288.99877 & Fe I   \\
61.1831 & Kr IV  & 128.395   & Ar VI  &	 294.5106  & He I   \\
62.165  & Ar VII & 130.387   & Ar VI  &	 297.9860  & W I    \\
63.72881& Ar III & 132.3847  & W IV   &	 300.0     & Fe I   \\
66.200  & Ar IX  & 134.3710  & Ar III &	 303.01481 & Fe I   \\
67.0948 & Ar II  & 135.6086  & W III  &	 309.0088  & W I    \\
67.897  & Kr IV  & 137.9670  & Al III &	 314.0146  & W I    \\
68.8915 & Kr VII & 139.3886  & W III  &	 321.01863 & Fe I   \\
69.9812 & Kr IV  & 142.3889  & W V    &	 324.993   & Mo I   \\
72.394  & Ar VII & 145.8088  & W IV   &	 334.97241 & Fe I   \\
73.0    & Fe III & 149.1151  & W III  &	 337.997   & Mo I   \\
74.011  & Ar VIII& 154.3841  & W IV   &	 343.003   & Ar III \\
76.8132 & Kr III & 155.982   & Ar IV  &	 348.983   & O IV   \\
78.216  & Kr IV  & 157.592   & La IV  &	 350.98614 & Fe I   \\
80.109  & Ar IV  & 158.8466  & W III  &	 356.006   & W I    \\
82.598  & Kr VI  & 160.3074  & Ar II  &	 357.995   & Kr III \\
83.417  & Kr VI  & 162.790   & Kr IV  &	 364.0134  & W I    \\
84.2805 & Ar I   & 165.5639  & Cr V   &	 365.97333 & Fe I   \\
85.0154 & Kr I   & 167.211   & Kr IV  &	 367.99131 & Fe I   \\
86.832  & Ar IV  & 169.028   & Cr III &	 372.999   & W I    \\
87.792  & Ar VII & 178.5729  & W III  &	 377.03012 & Fe I   \\
89.394  & Kr IV  & 183.001   & Ni III &	 380.9848  & W I    \\
90.0313 & Kr I   & 188.77646 & Fe I   &	 384.99664 & Fe I   \\
91.9342 & H I    & 192.0378  & W III  &	 386.9928  & W I    \\
93.0749 & H I    & 197.39162 & Fe I   &	 393.0231  & W I    \\
94.9742 & H I    & 206.3848  & W IV   &	 397.98498 & W II   \\
96.4075 & Kr IV  & 213.20171 & Fe I   &	 403.9938  & Fe I   \\
97.2541 & H I    & 217.80806 & Fe I   &  407.98377 & Fe I   \\
98.03   & Ar IV  & 223.008   & Cu I   & &\\		
99.004  & Ar IV  & 228.998   & Ni I   & &\\		
100.1883& W IV   & 232.003   & Ni I   & &\\ 		
102.5948& W IV   & 241.9985  & W II   & &\\
\noalign{\smallskip}\hline\noalign{\smallskip}
\end{tabular}}
\end{table*}

\newpage
\begin{table*}[!ht]
\centering
\caption{Validated spectra lines in the wavelength range from $138$ to $160$ nm. The lines are depicted in Fig. \ref{s2} for a gas flow of $1200$ and $2000$ sccm. Data are taken from the NIST database \citep{nist}.}
\label{tab:2}  
\scalebox{0.8}{ \begin{tabular}{ll|ll|ll|ll}
\noalign{\smallskip}\hline\noalign{\smallskip}
Wavelength [nm] & Element & Wavelength [nm] & Element & Wavelength [nm] & Element & Wavelength [nm] & Element   \\
\noalign{\smallskip}\hline\noalign{\smallskip}
138.0723 & Ar II & 144.766 & Ar V & 152.411 & Kr IV & 157.5375 & Kr II \\
138.2228 & Ar II  & 144.835 & Ar IV & 152.515 & Kr IV & 157.566 & Kr IV\\
138.235 & Kr V  & 145.1879 & Ar II & 152.5486 & Kr II & 157.5815 & Ar II\\
138.2765 & Ar II & 145.226 & Kr V  & 152.566 & Kr IV & 157.594 & Kr IV \\
138.386 & Kr V  & 145.348 & Kr III & 152.626 & Ar IV & 157.6155 & Kr II \\
138.459 & Kr V  &  145.544 & Ar IV & 152.795 & Ar IV & 157.65915 & Ar III\\
138.511 & Kr IV  & 145.5484 & Ar II & 152.818 & Ar IV & 157.6897 & Ar II \\
138.681 & Kr IV  & 145.630 & Kr IV & 153.250 & Kr III & 157.8812 & Ar II \\
138.900 & Kr IV  & 145.730 & Kr IV & 153.286 & Kr IV & 157.9513 & Kr II \\
138.996 & Kr IV  & 145.923 & Ar IV & 153.341 & Kr IV & 157.9731 & Kr II \\
139.212 & Ar XI  & 145.948 & Kr IV & 153.553 & Kr IV & 157.974 & Kr IV \\
139.263 & Kr V  & 145.9875 & Ar II & 153.600 & Ar IV & 158.0260 & Ar III  \\
139.361 & Kr V  & 146.00973 & Ar III & 153.668 & Ar IV & 158.0768 & Ar II \\
139.564 & Ar IV  & 146.02487 & Ar II & 153.855 & Kr V  & 158.0960 & Ar II\\
139.6231 & Ar II  & 146.099 & Ar IV & 153.9075 & Kr II & 158.198 & Kr IV \\
140.089 & Kr IV  & 146.109 & Kr IV & 154.007 & Kr V & 158.248 & Kr III \\
140.168 & Kr III  & 146.265 & Kr XXIII & 154.026 & Ar IV & 158.30377 & Ar III \\
140.181 & Kr IV  & 146.3 & Ar VIII & 154.2540 & Ar III & 158.383 & Ar II\\
140.220 & Kr V  & 146.3155 & Ar II & 154.291 & Kr V & 158.4563 & Kr II\\
140.288 & Kr III  & 146.4072 & Kr II & 154.4177 & Ar II & 158.601 & Kr III \\
140.750 & Ar IV  & 146.4176 & Ar II & 154.4711 & Ar II & 158.6093 & Kr II \\
140.930 & Ar IV  & 146.5153 & Ar II & 154.508 & Kr IV & 158.6170 & Kr II \\
141.013 & Ar IV  & 146.553 & Kr IV & 154.630 & Kr IV & 158.6256 & Ar II \\
141.157 & Kr IV  & 146.55506 & Ar III & 164.666 & Ar IV & 158.6330 & Ar III \\
141.235 & Kr III  & 146.57036 & Ar III & 154.7354 & Ar II & 158.66206 & Ar III\\
141.3894 & Kr II  & 146.600 & Kr IV & 155.452 & Ar IV & 158.6621 & Kr II\\
141.397 & Ar V  & 146.614 & Kr IV  & 155.539 & Kr V & 158.8740 & Ar III\\
141.614 & Kr V  & 146.6460 & Kr II & 155.6220 & Ar III & 158.9384 & Kr II \\
141.689 & Kr IV  &  146.6524 & Ar II & 155.6630 & Ar III & 158.9463 & Ar II\\
141.959 & Ar IV  &  146.78533 & Ar III & 155.7302 & Ar II & 158.987 & Kr IV\\
142.060 & Ar VI  &  146.8006 & Ar III & 155.851 & Kr IV & 159.0229 & Ar II\\
142.070 & Kr III  & 146.8021 & Kr II & 155.8802 & Kr III & 159.032 & Kr IV \\
142.171 & Kr IV  & 147.204 & Ar IV & 155.9072 & Ar II & 159.160 & Kr III \\
142.2000 & Ar III  & 147.2594 & Ar II & 155.982 & Ar IV & 159.1933 & Ar II \\
142.251 & Ar VI  & 147.448 & Ar IV & 156.0184 & Ar II & 159.2565 & Kr II\\
142.2512 & Kr II  & 147.4537 & Ar II & 156.193 & Kr IV & 159.3581 & Ar II \\
142.3553 & Kr III  & 147.717 & Kr IV & 156.2441 & Ar II & 159.386 & Kr IV\\
142.429 & Kr IV  & 147.915 & Ar IV & 156.285 & Kr III & 159.3946 & Kr II \\
142.497 & Kr V  &  148.136 & Kr IV & 156.3036 & Ar II & 159.4787 & Ar II\\
142.575 & Kr III  & 148.160 & Ar IV & 156.5377 & Ar II & 159.4895 & Kr II \\
142.619 & Ar IV & 148.3429 & Kr III & 156.603 & Kr V & 159.5734 & Ar II \\
142.777 & Kr III  & 148.601 & Ar IV & 156.6812 & Ar II & 159.6141 & Ar II \\
142.965 & Kr IV  & 148.628 & Kr IV & 156.7987 & Ar II & 159.6210 & Ar III \\
142.984 & Kr V  & 148.952 & Kr IV & 156.8050 & Kr II & 159.641 & Kr IV \\
143.068 & Kr IV  & 149.018 & Kr IV & 156.8690 & Ar III & 159.667 & Kr IV \\
143.378 & Kr V  & 149.0928 & Kr II & 156.891 & Kr V & 159.8082 & Kr II\\
143.4070 & Ar III  & 149.1104 & Kr II & 156.9135 & Kr II & 159.8561 & Ar II \\
143.5085 & Kr II  & 149.532 & Ar IV & 156.982 & Kr IV & 159.8724 & Ar II \\
143.557 & Ar IV  & 149.5769 & Kr II & 156.9886 & Kr III & 159.8872 & Ar II \\
143.5676 & Kr II  & 149.781 & Ar IV  & 157.017 & Ar IV & 159.9125 & Ar II\\
143.620 & Kr V  & 149.828 & Kr V & 157.038 & Kr IV & 159.9492 & Kr II \\
143.648 & Kr IV  & 149.850 & Kr III & 157.1390 & Ar II & 159.9597 & Ar II \\
143.7020 & Ar III  & 149.964 & Kr V & 157.1876 & Kr II & 159.98 & Kr V \\
143.7170 & Ar III  & 150.0740 & Ar III & 157.1920 & Ar III & 159.982 & Kr IV \\
143.912 & Kr IV  & 150.158 & Kr IV & 157.23340 & Ar III & & \\
144.007 & Ar III  & 150.290 & Kr IV & 157.2340 & Kr II  & &\\
144.0210 & Ar III  & 150.591 & Kr III & 157.3050 & Ar III & & \\
144.2440 & Ar III  & 150.880 & Ar IV & 157.318 & Kr III & & \\
144.343 & Kr IV  & 150.984 & Ar IV & 157.3404 & Kr II  & & \\
144.4 & Ar VIII  & 151.351 & Kr IV & 157.4103 & Kr II & & \\
144.4343 & Kr II  & 151.4585 & Kr II & 157.4340 & Kr II & & \\
144.528 & Kr IV  & 151.605 & Ar IV & 157.4402 & Ar II & &\\
144.563 & Kr IV  & 151.685 & Kr IV & 157.4733 & Kr II  & &\\
144.750 & Kr III  & 152.162 & Kr IV & 157.498 & Kr IV & &\\
144.762 & Kr IV  & 152.371 & Kr IV & 157.4992 & Ar II & &\\
\noalign{\smallskip}\hline\noalign{\smallskip}
\end{tabular}}
\end{table*}

\newpage

\end{document}